\documentclass[aps, prd, 10pt, twocolumn, superscriptaddress,noshowpacs, preprintnumbers
,nofootinbib,floatfix]{revtex4-2}

\usepackage[dvipsnames]{xcolor}
\usepackage[colorlinks=true,breaklinks=true]{hyperref}
\hypersetup{allcolors=[rgb]{0.0 0.0 0.70},linkcolor=[rgb]{0.75 0.05 0.05}}
\usepackage{orcidlink}
\usepackage{microtype}

\usepackage{amsmath}
\usepackage{amsfonts}
\usepackage{amssymb}
\usepackage{mathtools}
\usepackage{multirow}
\usepackage{bm}
\usepackage[utf8]{inputenc}

\usepackage{pifont}

\begin{document}

\title{Detection horizon for the neutrino burst from the stellar helium flash}

\author{Pablo Mart\'inez-Mirav\'e \orcidlink{0000-0001-8649-0546}}
\email{pablo.mirave@nbi.ku.dk}
\affiliation{Niels Bohr International Academy and DARK, Niels Bohr Institute, University of Copenhagen,\\
Blegdamsvej 17, 2100, Copenhagen, Denmark}

\author{Irene Tamborra \orcidlink{0000-0001-7449-104X}}
\email{tamborra@nbi.ku.dk}
\affiliation{Niels Bohr International Academy and DARK, Niels Bohr Institute, University of Copenhagen,\\
Blegdamsvej 17, 2100, Copenhagen, Denmark}

\author{Georg Raffelt \orcidlink{0000-0002-0199-9560}}
\email{raffelt@mpp.mpg.de}
\affiliation{Max-Planck-Institut f{\"u}r Physik, Boltzmannstrasse 8, 85748 Garching, Germany}

\begin{abstract}
Low-mass stars ($M\lesssim 2\,M_\odot$) ignite helium under degenerate conditions, eventually causing a nuclear run-away---the helium flash. The alpha-capture process on $^{14}$N produces a large amount of $^{18}$F, whose subsequent decay spawns an intense $\nu_e$ burst (with average energy of $0.38$~MeV)  lasting about a day. We show that, in addition, a strong $1.7$~MeV neutrino line is generated by electron capture on $^{18}$F. Detection is hindered by large backgrounds in state-of-the-art neutrino observatories, such as JUNO. In next-generation facilities, such as the Jinping neutrino experiment, the horizon for a detection with a local significance of $3 \sigma$ would be extended to almost $3$~pc. Although helium flashes occur a few times per year in our Galaxy, there are no stellar candidates  approaching the tip of the red giant branch within $10$~pc. Hence, to date,  asteroseismology remains the most promising tool for probing the most energetic thermonuclear event in the life of a low-mass star.
\end{abstract}

\maketitle

\section{Introduction}
Neutrinos with energies in the keV--MeV energy range are copiously produced throughout the lives of stars \cite{Martinez-Mirave:2025dae, Farag:2020nll, Farag:2023xid}, yet to date only the Sun has been observed as a stellar source \cite{Davis:1968cp, Antonelli:2012qu, Haxton:2012wfz}, with the notable exception of supernova SN~1987A \cite{Mirizzi:2015eza, Fiorillo:2023frv, Raffelt:2025wty}. Neutrino observations have been instrumental in shaping our understanding of the closest star, while simultaneously advancing neutrino physics itself \cite{Acharya:2024lke, Gann:2021ndb}. The next stellar neutrinos to be detected are likely those of the diffuse supernova neutrino background, originating from all past collapses of massive stars, or even those from the next Galactic supernova \cite{2023PJAB...99..460A,Tamborra:2024fcd,Horiuchi:2018ofe}. For a very nearby stellar core collapse, neutrinos from the final nuclear burning stages of the progenitor may even be observable~\cite{Patton:2017neq,Mukhopadhyay:2020ubs}. For all other stars, the signal will likely remain hidden under the overwhelming solar foreground that dominates the low-energy neutrino sky~\cite{Vitagliano:2019yzm}, although stellar neutrinos from the inner Galaxy may eventually become observable~\cite{Martinez-Mirave:2025dae}.

It is, therefore, intriguing that the transient episode linked to He ignition in low-mass stars, known as the He flash, engenders an intense burst of nuclear neutrinos. The detection prospects of  the latter were first explored by Serenelli and Fukugita \cite{Serenelli:2005nh} and are re-examined here. A low-mass star ($M\lesssim 2 M_\odot$) develops a degenerate He core that grows to a mass of about $0.5\,M_\odot$ before He ignites at the tip of the red-giant branch (TRGB). The core consists primarily of $^4$He and heavier elements left behind by CNO burning, which appear almost entirely in the form of $^{14}$N, the bottleneck of the CNO cycle. Concurrently to helium ignition via the \hbox{triple-$\alpha$} ($3 \alpha$) reaction, $^{14}$N is converted to $^{18}$F through $\alpha$-capture. The subsequent decay $^{18}{\rm F}\to {}^{18}{\rm O}+e^++\nu_e$, with a half-life of $1.83$~h, produces a copious $\nu_e$ burst with an average energy of $0.38$~MeV. This phenomenon was called nitrogen flash in Ref.~\cite{Serenelli:2005nh}, whereas we  refer to the $^{18}{\rm F}$ neutrinos.

The runaway thermonuclear burning lasts up to a few days, until the core has expanded enough to lift electron degeneracy and  allow for quiescent, self-regulated nuclear burning. The He flash is the most energetic thermonuclear event in the life of a low-mass star~\cite{Sweigart:1978}. At  peak, the associated energy release can be as large as  $10^{10}\,L_\odot$ (where $L_\odot=3.83\times10^{33}~{\rm erg}~{\rm s}^{-1}$ is the solar luminosity) and the   $^{18}$F neutrino production reaches $10^{47}~{\rm s}^{-1}$. Although it occurs  for a brief period only, the latter should be compared with $1.8\times10^{38}~{\rm s}^{-1}$ of the Sun. All nuclear reactions in the stars of the Galaxy produce approximately $10^{49}~{\rm s}^{-1}$ neutrinos \cite{Martinez-Mirave:2025dae}.

Despite its dramatic energetics, the  He flash has no directly visible signature. Of course, low-mass red giants brighten only until they reach the TRGB and then become helium-burning stars on the horizontal branch. However, the associated dimming takes a long time---one cannot observe red giants disappearing at the TRGB. Asteroseismology is perhaps the most promising probe: convection driven by the He flash can excite gravity modes that propagate to the stellar surface, inducing photometric modulations with periods of $10^3$--$10^4$~s~\cite{Bildsten:2012, Bertolami:2020, Capelo:2023pbf}. 

Neutrinos could be key probes of this stage of stellar evolution \cite{Serenelli:2005nh}, although a practical detection was considered  unlikely. In this paper, building on our previous study of neutrino emission from all stars in our Galaxy~\cite{Martinez-Mirave:2025dae}, we re-examine this question in the context of modern neutrino observatories. In addition to emission via $\beta^+$ decay, electron capture (EC) on $^{18}$F, i.e., $e^-+{}^{18}{\rm F}\to {}^{18}{\rm O}+\nu_e$ ($1.7$~MeV), produces a monochromatic line that could offer additional background discrimination. Unfortunately, the sensitivity in current and next-generation neutrino observatories is insufficient given that there are no He-flash candidates   within about $10$~pc from Earth.

This paper is arranged as follows. The modelling of the He flash is outlined in Sec.~\ref{sec:nu_model}, whereas neutrino production from $^{18}$F during the He flash is presented in Sec.~\ref{sec:nuemission}. Section~\ref{sec:detection} explores the detection prospects in idealized neutrino observatories, and Sec.~\ref{sec:conclusions} summarizes our findings. Appendix~\ref{sec:F18-decay} provides details about the $^{18}$F decay and the relative strength of the $\beta^+$ and EC channels. Appendix~\ref{sec:He-flash-rate} estimates the He flash rate in the Galaxy.

\section{The core helium flash}
\label{sec:nu_model}

In this section, we introduce a reference stellar model to investigate the neutrino burst associated with the He flash. We use  spherically symmetric stellar models with zero-age main-sequence (ZAMS) masses of 1--$2\,M_\odot$ and solar metallicity, evolved with the \texttt{Modules for Experiments in Stellar Astrophysics} (\texttt{MESA})  code.

\subsection{Features of the helium flash}

Low-mass stars on the red-giant branch have degenerate helium cores with well-understood properties \hbox{\cite{Sweigart:1978, Salaris:2002xf, Kippenhahn:2012qhp}}. Shell hydrogen burning continuously increases the mass of the core, which contracts following the inverse mass-radius relation of degenerate stars, and  heats up by gravitational energy release. Neutrino emission by plasmon decay tends to cool the central core region, causing a temperature maximum significantly away from the center, and it is this shell where He eventually ignites. This process was named He flash, given the initial uncertainties surrounding its physics and its potentially dynamical, even explosive, nature. However, state-of-the-art stellar models suggest that this transient event is far more benign. 

Multi-dimensional models have been instrumental to grasp the features of the He flash \cite{Dearborn:2005bb, Mocak:2008kf, Mocak:2008ht}; repeated subflashes (with much smaller energy release), initially observed in spherically symmetric models \cite{Thomas:1967, Serenelli:2005qs}, may not exist in three-dimensional stellar models. Although  the stars begin to dim when the core starts expanding, the overall readjustment is too slow to be directly observed. The huge amount of  energy release turns  into core expansion without immediate impact on stellar surface properties. An asteroseismological signature consists of the excitation of gravity waves and associated brightness \hbox{modulations~\cite{Bildsten:2012, Bertolami:2020, Capelo:2023pbf}}.

\subsection{Numerical stellar models}

In view of the exploratory nature of our work, we rely upon spherically symmetric stellar models. They are evolved through the He flash with  the publicly available code \texttt{MESA} 24.08.1 \cite{Paxton:2011, Paxton:2013pj, Paxton:2015jva, Paxton:2019, MESA:2022zpy}. We adapt existing models from the \texttt{MESA} Test Suite following the procedure outlined in Ref.~\cite{Martinez-Mirave:2025dae}. We adopt a mixing-length treatment of convection~\cite{Henyey:1965, Paxton:2011}, and parametrize the diffusion coefficient as in Ref.~\cite{Herwig:2000sq} to account for convective overshooting. We rely on the built-in nuclear reaction network \texttt{mesa\_49}, which  captures all  nuclear and weak processes relevant  for stars in the mass range of interest. For additional details on the convection treatment and the nuclear and weak reaction rates, see Ref.~\cite{Martinez-Mirave:2025dae}. We consider three  models with ZAMS mass of $1$, $1.8$, and $2\,M_\odot$ and solar \hbox{metallicity}. Specifically, the metal mass fraction is $Z=0.0182$, and of this, $0.0125$ are the CNO elements, whereas the initial He mass fraction is $Y=0.2736$. 

\begin{figure}
    \centering
    \includegraphics[width=\linewidth]{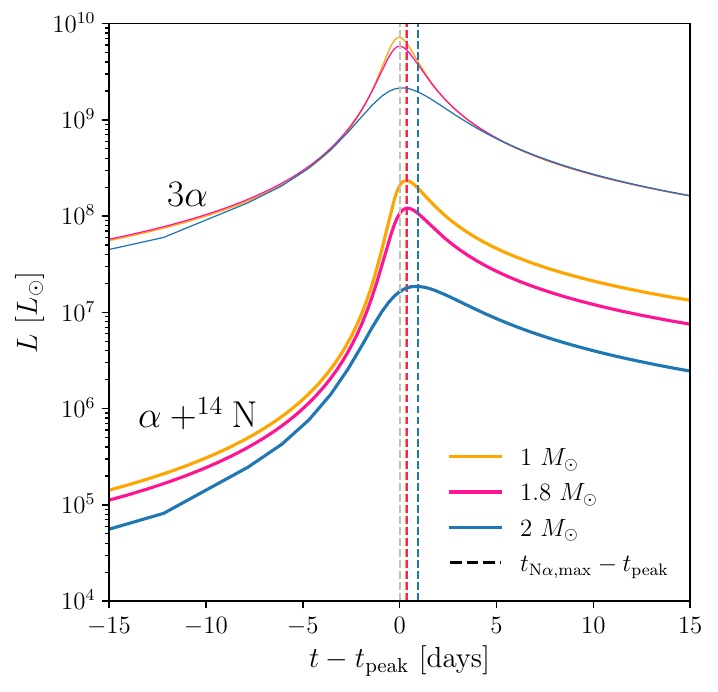}
    \vskip-4pt
    \caption{Evolution of the rate of energy release during the He flash of three models with ZAMS mass of $1$, $1.8$, and $2 \,M_\odot$. The upper (thin) lines refer to the $3\alpha$ reaction, the lower (thick) lines to $^{14}{\rm N}(\alpha,\gamma)$, leading to ${}^{18}{\rm F}$ production. Time is relative to $t_{\rm peak}$, i.e., the instant of maximum $L_{3\alpha}$. The vertical lines show the time offset of the maximum of $L_{{\rm N}\alpha}$ relative to $t_{\rm peak}$.
    }
    \label{fig:luminosity}
    \vskip-4pt
\end{figure}

Figure~\ref{fig:luminosity} shows the evolution of the energy-generation rate around the He flash. In particular, we focus on the energy-production rate $L_{3\alpha}$ due to the triple-$\alpha$ reaction (upper curves) and $L_{{\rm N}\alpha}$ due to $\alpha$ capture on ${}^{14}$N (lower curves). Time is measured with respect to  $t_{\rm peak}$, the instant of maximum $L_{3\alpha}$. The peak of $L_{{\rm N}\alpha}$ is slightly delayed as one can infer from the vertical lines. For the $1\,M_\odot$ model, $L_{3\alpha}$ increases by about two orders of magnitude starting from two weeks before the peak and then decreases as the He burning region expands. $L_{{\rm N}\alpha}$ increases even more dramatically in this period, while it tracks $L_{3\alpha}$ after the peak.

The evolution of the energy release is similar for the  $1\,M_\odot$ and  $1.8\,M_\odot$ models, as well as several ones with intermediate mass that we have run (not shown here). For the $2\,M_\odot$ model, the  evolution of the energy release  is  flatter. For somewhat larger masses, there is no longer a He flash, as He ignites before developing a degenerate core. Moreover, our simulations show that the ignition radius is ever larger for the models with smaller masses, whereas for $M\agt 2\,M_\odot$, the temperature maximum is at the center (results not shown here).

The mixing-length treatment in such spherically symmetric models generally predicts convection to develop from the radius of He ignition outwards~\cite{Mocak:2008}. As a consequence, these models tend to predict a succession of He subflashes occurring closer to the center. This is indeed the case in our models. However, since this feature is questioned by multidimensional simulations (cf., e.g., Ref.~\cite{Dearborn:2005bb}), we focus on the primary He flash.

Our  model with $1\,M_\odot$ is representative for the He flash in stars with ZAMS mass not too close to the limiting value of around $2\,M_\odot$, beyond which there is no flash. Therefore, in the following, we restrict ourselves to the $1\,M_\odot$ case, where we find He to ignite when $M_{\rm core}=0.48\,M_\odot$. The ignition point is at a mass coordinate from the center of $m=0.17\,M_\odot$, where shortly before He ignition, $\rho\simeq5\times10^5~{\rm g}~{\rm cm}^{-3}$ and $T=10^8~{\rm K}$.

\begin{figure}
    \centering
    \vskip-4pt
   \includegraphics[width=\linewidth]{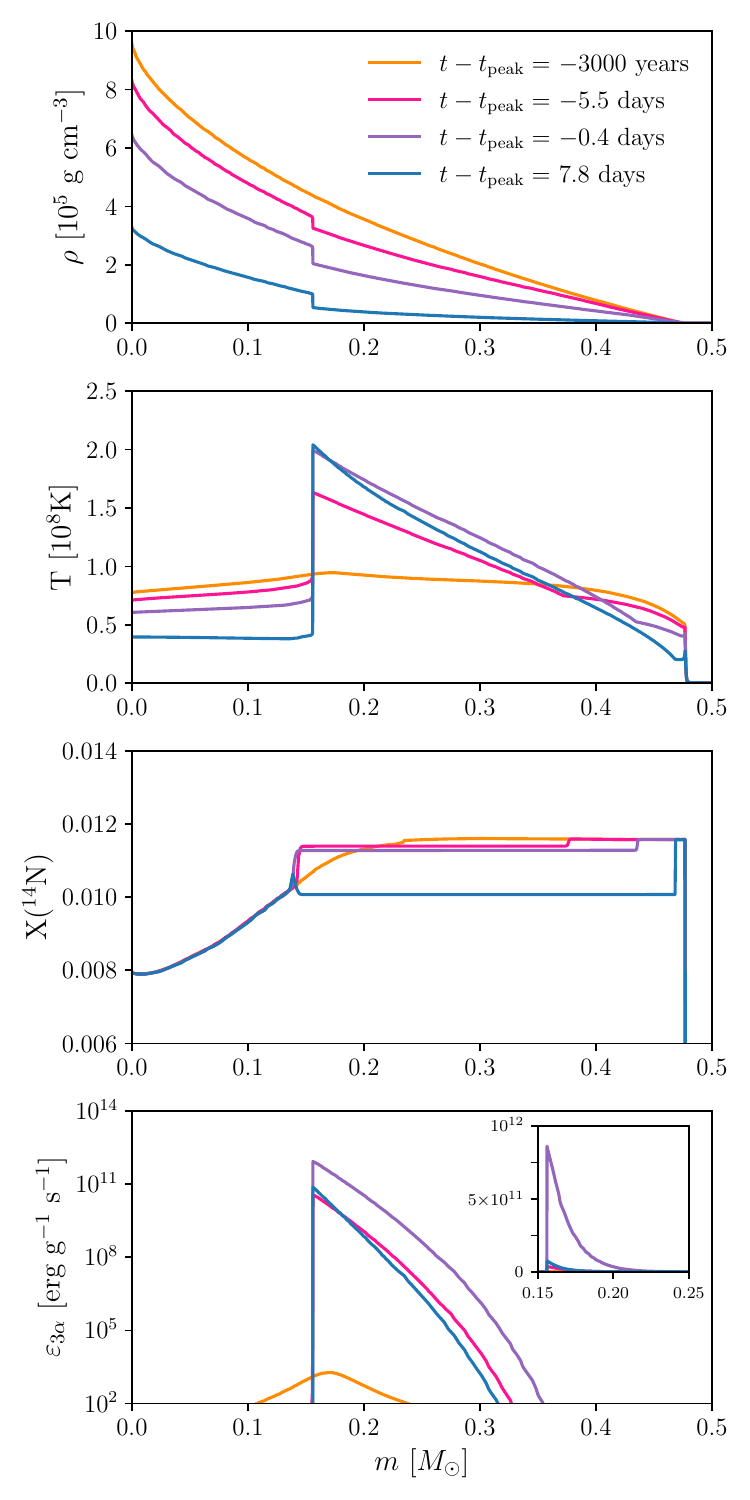}
    \vskip-4pt
    \caption{Core profiles of the $1\,M_\odot$ model for selected times before and after He ignition ($t-t_{\rm{peak}} = -3000$~years, $-5.5$~days, $-0.4$~days, and $7.8$~days). From top to bottom, we show the baryon density $\rho$, medium temperature $T$, the $^{14}$N abundance by mass, and the $3\alpha$ energy generation rate per unit mass. The inset of the bottom panel displays the energy generation rate in the proximity of $t_{\rm{peak}}$.
    }
    \label{fig:profiles}
    \vskip-4pt
\end{figure}

\begin{figure}
    \centering
   \includegraphics[width=\linewidth]{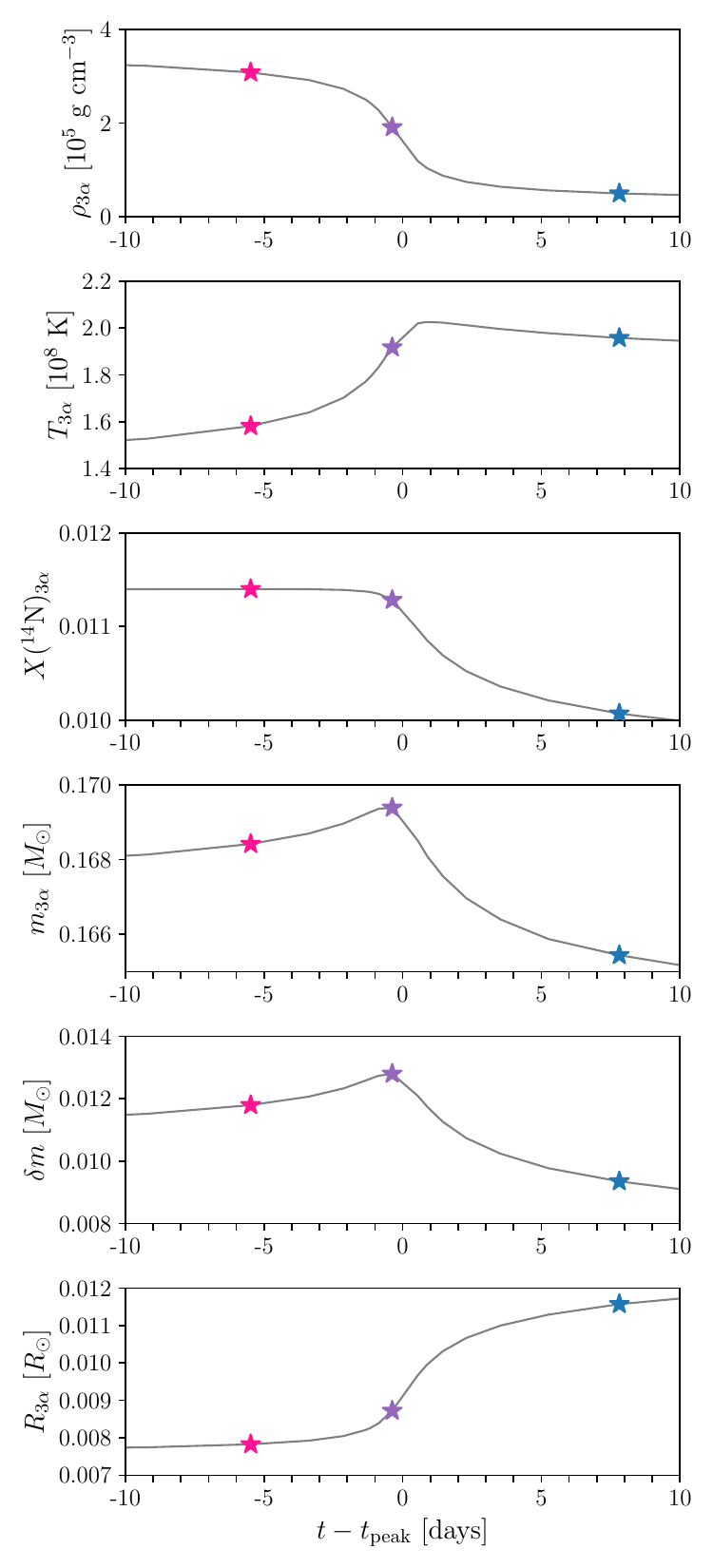}
    \caption{Evolution of several physical quantities in the He-burning region, defined through Eq.~\eqref{eq:Qav}. From top to bottom, we show $\rho$, $T$, $X_{^{14}{\rm N}}$, the mass coordinate ($m$), the width of the relevant mass region ($\delta m$), and the radial coordinate $R$. The colored asterisks mark  the snapshots shown in Fig.~\ref{fig:profiles}, except for the pre-ignition profile at $-3000$~years.}
    \label{fig:evolution-peak}
\end{figure}

Figure~\ref{fig:profiles} shows the profiles of several physical quantities as  functions of the mass coordinate $m$ for several time snapshots around He ignition, where time is defined with respect to $t_{\rm peak}$. The orange line ($t - t_{\rm peak}=-3000$~years) represents the conditions before He begins to ignite and the core starts expanding. As displayed in the second panel, the orange line shows a temperature maximum at $m=0.17\,M_\odot$, where the 3$\alpha$ reaction runs away. 

The third panel of Fig.~\ref{fig:profiles} shows the $^{14}$N abundance
profile, which increases from $X(^{14}{\rm N})=0.8\%$ by mass at the center $1.18\%$ in the outer core, and drops to the initial value of $0.14\%$ beyond the hydrogen burning shell, which is at $m=0.48\,M_\odot$. Notice that most of the CNO elements are initially in the form of $^{16}$O and $^{12}$C, whereas $^{14}$N contributes around an order of magnitude less \cite{Asplund:2009fu}. Our $1\,M_\odot$ star, like the Sun, initially burns hydrogen mostly through the pp chains, when the CNO cycle does not reach equilibrium. The later shell burning, via the CNO cycle, puts nearly all of the CNO elements into $^{14}$N. 

The profiles extracted a few days around  He ignition show how the temperature shoots up at the point of ignition (second panel  of Fig.~\ref{fig:profiles}), while the entire core expands. In these spherically symmetric models, convection occurs only outside of the He burning shell, which is very narrow; in the bottom panel, we see a steep decrease in the energy generation rate outside of the immediate burning shell. An energy generation rate per unit mass near maximum of around $8\times10^{11}~{\rm erg}~{\rm g}^{-1}~{\rm s}^{-1}$ and a peak value for $L_{3\alpha}$ of around $8\times10^{9}\,L_\odot$ corresponds to a burning mass of around $0.02\,M_\odot$. Fresh fuel, and notably $^{14}$N, is dredged down to the burning region by convection. We can see the corresponding $^{14}$N depletion. In other words, a significant fraction of the total amount of $^{14}$N disappears within a week after ignition, going from $4.9\times 10^{53}$ nuclei before He ignition to $4.4\times 10^{53}$ one week later.

Figure~\ref{fig:evolution-peak} shows the evolution of several characteristic quantities in the region of He burning. We define them as averages over the core, weighted with $L_{3\alpha}$ itself, i.e., for a quantity $Q(m)$ in the form
\begin{align}\label{eq:Qav}
    \langle Q\rangle_{\rm 3\alpha} = \frac{\int Q(m)\, L_{3\alpha}(m){\rm d}m}{\int  L_{3\alpha}(m){\rm d}m}\, .
\end{align}
These averages provide a first impression of the conditions in the He-burning and neutrino-producing region. We see how this region both heats up and decreases in density. The mass coordinate for the narrow region of He burning, $\langle m\rangle_{3\alpha}$, essentially remains at the original ignition point, but moves to a larger geometric radius, $R_{3\alpha}$. The width of the mass region, $\delta m$, where He burning takes place is also shown, defined here as the standard deviation $\delta m = \sqrt{\langle m^2\rangle_{3\alpha} - \langle m\rangle^2_{3\alpha}}$.

\section{Neutrino emission from  \texorpdfstring{\boldmath{$^{18}$}}{}F}
\label{sec:nuemission}

In this section, we explore the features of neutrino emission from $^{18}$F during the He flash. We also outline the optimal window for neutrino detection and the neutrino flux expected at Earth.

\subsection{Neutrino emission}
\label{sec:nuemission-profile}

    In a low-mass star near the TRGB, neutrinos are produced  by different processes. The main source of energy release  is the CNO cycle in the hydrogen-burning shell surrounding the degenerate He core. In our $1\,M_\odot$ model, the bolometric luminosity at the TRGB is $3.5\times10^3\,L_\odot$, with a corresponding nuclear neutrino generation rate of roughly $5\times 10^{41}~{\rm s}^{-1}$. This region also produces  thermal neutrinos with keV energies, similar to  the  thermal  flux of solar neutrinos \cite{Vitagliano:2017odj}. 

Within the core, thermal neutrinos are produced primarily by plasmon decay and have energies of $\mathcal{O}(30)$~keV, corresponding to a core temperature of around $10$~keV. This channel of neutrino cooling is most effective in the densest region and thus causes a local  minimum of $T$ at the center and, consequently, off-center He ignition.

All of these fluxes are negligible compared with the $^{18}$F neutrino burst. Most of the original CNO elements, a mass fraction of $0.0182$ in our $1\,M_\odot$ model, piles up as~$^{14}$N, the bottleneck of the CNO cycle. During helium burning, $^{14}$N converts to $^{18}$F by $\alpha$ capture. The $^{18}$F population then decays with a half-life of $1.83$~hours~\cite{GarciaTorano:2010} or undergoes EC
\begin{subequations}
  \begin{align}
     &{}^{18}{\rm F} \, \to {}^{18}{\rm O} \, + e^+ \, + \, \nu_e \quad (\beta^+\, {\rm decay})\, ,\\
     &{}^{18}{\rm F} \, + e^- \, \to {}^{18}{\rm O} \, + \, \nu_e \quad ({\rm EC})\, .
\end{align}
\end{subequations}
The reaction chain ends with $^{18}{\rm O}(\alpha,\gamma){}^{22}{\rm Ne}$. Details about the $^{18}$F processes are provided in Appendix~\ref{sec:F18-decay}. 

The maximum $\nu_e$ energy from $\beta^+$ decay is $0.634$~MeV, with an average of $0.38$~MeV. EC produces monoenergetic neutrinos with energy of $1.669$~MeV, a channel which contributes around $3\%$ to the laboratory decay width. However, in the dense environment of He ignition with an initial density of around $5\times10^5~{\rm g}~{\rm cm}^{-3}$, electrons are so dense that the EC rate is comparable with the  $\beta^+$ decay one, especially before core expansion by He ignition. The electrons are still essentially non-relativistic, with a kinetic energy of a few $10$~keV, which shifts and broadens the EC spectrum. It is still effectively a monochromatic line for which we adopt a nominal fixed energy of $1.68$~MeV (see Appendix~\ref{sec:F18-decay} for more details). As shown more explicitly later, this strong line source is better suited for  detection purposes  than the lower-energy continuum from $\beta^+$ decay, in view of the solar neutrino foreground. Therefore,  we  mostly focus on the EC neutrinos hereafter.

\begin{figure}
    \centering
    \includegraphics[width=\linewidth]{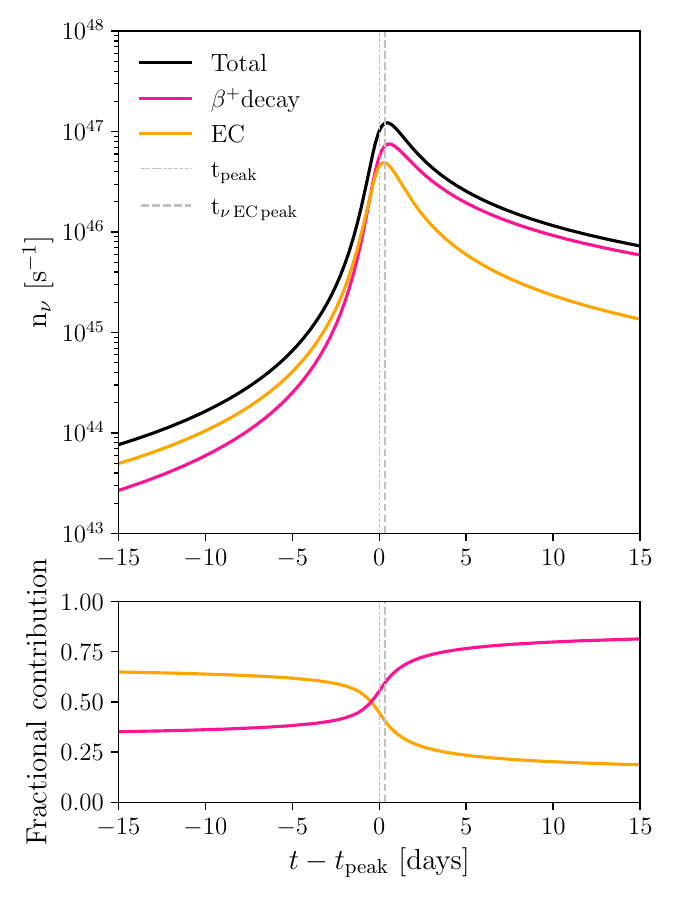}
    \caption{Evolution of $^{18}$F neutrino emission for our  $1\,M_\odot$ model. {\it Top panel:} Production by the two channels and their sum. The vertical lines indicate $t_{\rm peak}$ (corresponding to the maximum of  $L_{3\alpha}$) and the maximum of neutrino emission from EC.
    {\it Bottom panel:} Fractional contributions of $\beta^+$ decay and EC. At early times, EC dominates the neutrino emission rate; after the He flash, $\beta^+$ decay emission becomes the dominant process. }
    \label{fig:emissionbetaec}
\end{figure}

The nuclear network in \texttt{MESA} includes both of these processes, following the tabulation provided by Oda et al.\ \cite{Oda:1994} as a function of $\rho$ and $T$. As a standard output, both the neutrino number and energy production by the sum of the two channels is available, allowing us to extract the fractional contribution of both channels. For our reference model, the $^{18}$F neutrino production rates around the He flash are shown in Fig.~\ref{fig:emissionbetaec}.

The neutrino production rate essentially follows the  energy production rate of $^{14}N(\alpha, \gamma)$ in Fig.~\ref{fig:luminosity}. Its peak is also shifted relative to the one of $L_{3\alpha}$ by approximately $8$~hours, as indicated by the vertical lines. Before the peak, when electrons are still degenerate, the EC channel dominates, but drops in relative importance when the medium in the burning region begins to expand. The total number of emitted $\nu_e$ in the  time interval of $\pm15$~days around $t_{\rm peak}$ is $3.1\times 10^{52}$ from $\beta^+$ decay and $1.4\times10^{52}$ from EC, providing around half of the former, but still dominates in energy emission and detection rate due to the larger $E_\nu$. As mentioned earlier, hereafter we focus on the EC channel to explore its detection prospects.

\subsection{Time window for neutrino signal}
\label{sec:window}
The $^{18}$F burst is a transient signal at an unknown instant---it is not correlated with any electromagnetic nor gravitational-wave signature. Therefore, a detection strategy  requires investigating excess events in a low-energy neutrino detector above steady backgrounds, such as solar neutrinos, within a running time window of some chosen length $\tau$. What is the optimal value for $\tau$? If it is chosen too short,  signal is wasted; if it is  too long, the signal is diluted with more background.

\begin{figure}
    \centering
    \includegraphics[width=\linewidth]{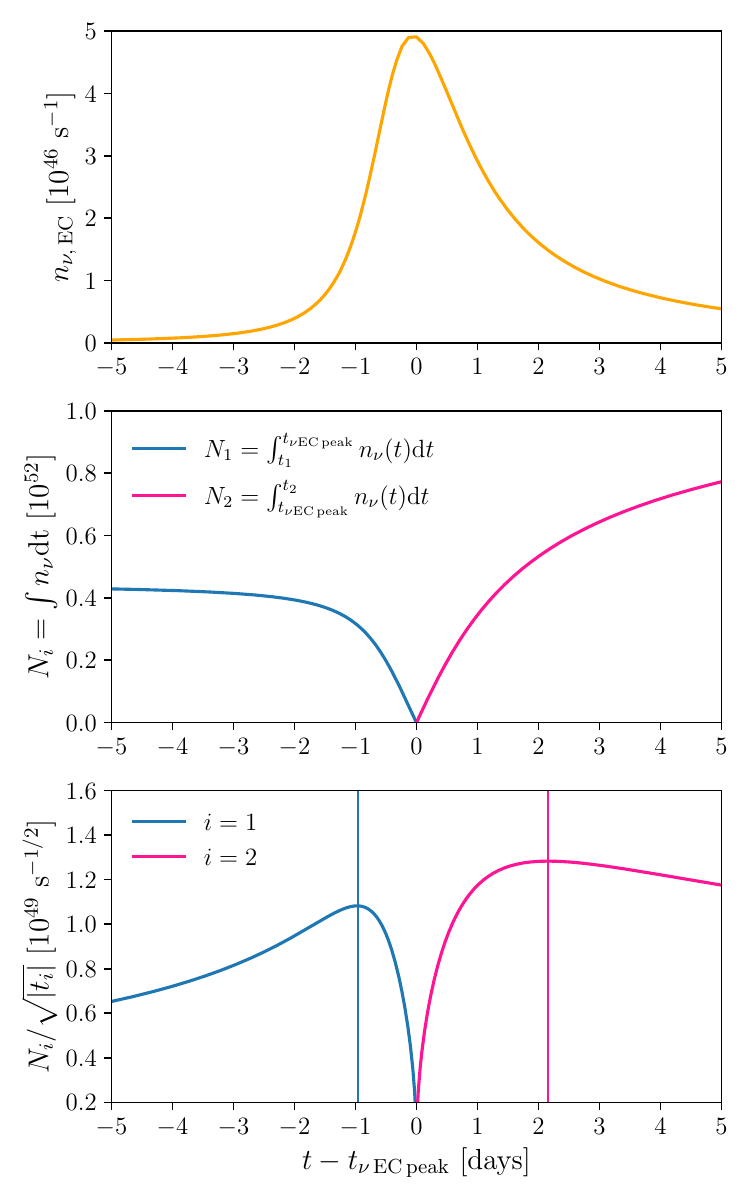}
    \caption{Evolution of $^{18}$F neutrino emission from EC for our   model with mass of $1\,M_\odot$. 
    {\it Top panel:} Emission rate, identical to the orange line of Fig.~\ref{fig:emissionbetaec}, now on a linear scale and centered on its maximum. {\it Middle panel:} Cumulative distribution relative to $t_{\nu\ \rm{EC\ peak}}$.
    {\it Bottom panel:} Cumulative rates divided by $\sqrt{|t|}$, with the maxima indicated by vertical lines. The maxima of $N_{1,2}/\sqrt{t_{1,2}}$ arise at $\tilde{t}_1=-0.96$~days and $\tilde{t}_2=2.15$~days, marked  by vertical lines. The optimal window for neutrino searches is given by $\tau=\tilde{t}_2-\tilde{t}_1=3.11$~days.
    }
    \label{fig:SB}
\end{figure}

Figure~\ref{fig:SB} (top panel) shows  the  EC neutrino signal on a linear scale, corresponding to the orange line in the upper panel of Fig.~\ref{fig:emissionbetaec}, now centered on $t_{\nu \,\rm{EC\, peak}}$. We observe that this flux spectrum resembles a skewed Lorentzian. Indeed, a surprisingly good analytic approximation, as a function of $\tilde{t} = t- t_{\nu \, \rm EC \, peak}$ in days, is
\begin{equation}
    n_\nu=4.91\times10^{46}~{\rm s}^{-1}\times
    \begin{cases}
        (1+0.7\,\tilde{t}\,^{3/2})^{-1}&\hbox{for $\tilde{t}>0$\, ,}\\[1ex]
        (1+1.9\,|\tilde{t}|^{5/2})^{-1}&\hbox{for $\tilde{t}<0$\, .}
    \end{cases}
\end{equation}
Within $-5<\tilde{t}<10$~days, this fit is accurate up to  a few percent and can serve as an approximate template for a refined detection strategy.

For a stellar neutrino emission rate $n_\nu(t)$, the number of detected events is proportional to the cumulative emission $N_\nu=\int_{t_1}^{t_2} n_\nu(t)\,{\rm d}t$, to be compared with $N_B=B(t_2-t_1)$, where $B$ is the background rate. The times $t_1$ and $t_2$ should be optimally chosen before and after the  maximum of $n_\nu$. One needs to compare background fluctuations with signal, 
i.e.,~$N_\nu$ with $\sqrt{N_B}$. Considering the one-sided intervals $N_1=\int_{{t}_1}^{t_{\nu\,\rm{EC\ peak}}} n_\nu(t)\,dt$ and $N_2=\int_{t_{\nu\, \rm{EC\ peak}}}^{{t}_2}n_\nu({t})\,d{t}$, these must be compared to  $\sqrt{B (t_{\nu\, \rm EC\, peak} - t_1)}$ and $\sqrt{B ({t}_2 - t_{\nu\, \rm EC\, peak})}$, respectively. Therefore, the optimal time interval for detection should be such that  the functions $N_{1,2}/\sqrt{|t_{1,2}-t_{\nu\, \rm EC\, peak}|}$ each have a maximum. The middle panel of Fig.~\ref{fig:SB} represents the cumulative neutrino number $N$.  In the bottom panel, the cumulative neutrino number is divided by $\sqrt{|t - t_{\nu\, \rm EC\, peak}|}$. The maxima of $N_{1,2}/\sqrt{|t_{1,2}-t_{\nu\, \rm EC\, peak}|}$ arise at $\tilde{t}_1=-0.96$~days and $\tilde{t}_2=2.15$~days, marked  by vertical lines in Fig.~\ref{fig:SB}. Hence, the optimal window for neutrino searches is given by  $\tau=\tilde{t}_2-\tilde{t}_1=3.11$~days.

\subsection{Flavor conversion}

All stellar neutrinos produced by these nuclear reactions  appear as $\nu_e$ and undergo nontrivial flavor evolution on the way to a detector. What exactly happens depends on the electron density at the point of production in the star and the density profile along the trajectory. The flavor evolution in our benchmark stellar model is adiabatic~\cite{Martinez-Mirave:2025dae}, hence only the neutrino energy and the density at the point of production matter. 

A Mikheyev-Smirnov-Wolfenstein (MSW)  crossing occurs when the electroweak potential $\sqrt{2}G_{\rm F} n_e$ equals the mass-induced energy difference $\Delta m_\nu^2/2E_\nu$. The solar and atmospheric mass differences are $\Delta m^2_{\rm L}=(8.6~{\rm meV})^2$ and  $\Delta m^2_{\rm H}=(50~{\rm meV})^2$~\cite{ParticleDataGroup:2024cfk}. For the EC line with $E_\nu=1.7$~MeV and assuming $2$ nucleons per electron, appropriate for He core conditions, this corresponds to mass densities for the L and H resonances of $\rho_{\rm L}=5.8\times10^2\,{\rm g}~{\rm cm}^{-3}$ and $\rho_{\rm H}=1.9\times10^4\,{\rm g}~{\rm cm}^{-3}$. For the average $\beta^+$ decay energy of $\langle E_\nu\rangle=0.38$~MeV, these numbers are a factor of $4.5$ larger, namely, $\rho_{\rm L}=2.6\times10^3\,{\rm g}~{\rm cm}^{-3}$ and $\rho_{\rm H}=8.6\times10^4\,{\rm g}~{\rm cm}^{-3}$---cf.~the density evolution in the He burning region  in Fig.~\ref{fig:evolution-peak}. Hence, we should expect that EC neutrinos  cross both MSW resonances adiabatically, but $\beta^+$ neutrinos with lower energy may only cross one of the  resonances; as a result, the flavor composition of neutrinos leaving the star varies as a function of  energy.

For solar neutrinos, where the corresponding mass densities are  somewhat smaller because of the large hydrogen abundance and correspondingly smaller number of nucleons per electron, flavor conversion for $E_\nu<1.7$~MeV effectively corresponds to near-vacuum oscillations, whereas adiabatic MSW conversion occurs for the higher-energy $^8$B neutrinos. The H resonance plays no role, therefore flavor conversion of solar neutrinos is almost entirely driven by the L sector. We account for  flavor conversion of solar neutrinos following Ref.~\cite{Martinez-Mirave:2024hfd}.

For normal mass ordering (NO), a state born as $\nu_e$ (a~propagation eigenstate) emerges at the stellar surface in the $\nu_3$ mass eigenstate, whereas in inverted ordering (IO), it emerges as $\nu_2$ \cite{Dighe:1999bi}. At detection, the states need to be projected onto the flavor basis. The original $\nu_e$ flux at production can then be discovered in the $\nu_e$ flavor with a survival probability $p$, and in one of the other flavors with  probability $1-p$. Ignoring possible Earth matter effects, the survival probability in the  adiabatic limit is~\cite{Dighe:1999bi}
\begin{subequations}
  \begin{eqnarray}
    \hbox to 2.5em{NO\hfil} p &=& \sin^2 \theta_{13}\simeq 0.022\, , \\
    \label{eq:IO-survival}
    \hbox to 2.5em{IO\hfil} p &=& \cos^2 \theta_{13}\sin^2\theta_{12}\simeq0.30\, .
  \end{eqnarray}
\end{subequations}
In NO, the original $\nu_e$ emerge as $\nu_3$, which has only a small $\nu_e$ component, hence the small survival probability. Of course, if detection occurs by elastic scattering on electrons, the non-$\nu_e$ flavors also have a sizeable neutral-current rate.

\subsection{Flux at Earth}

\begin{figure*}
    \centering
    \includegraphics[width=\linewidth]{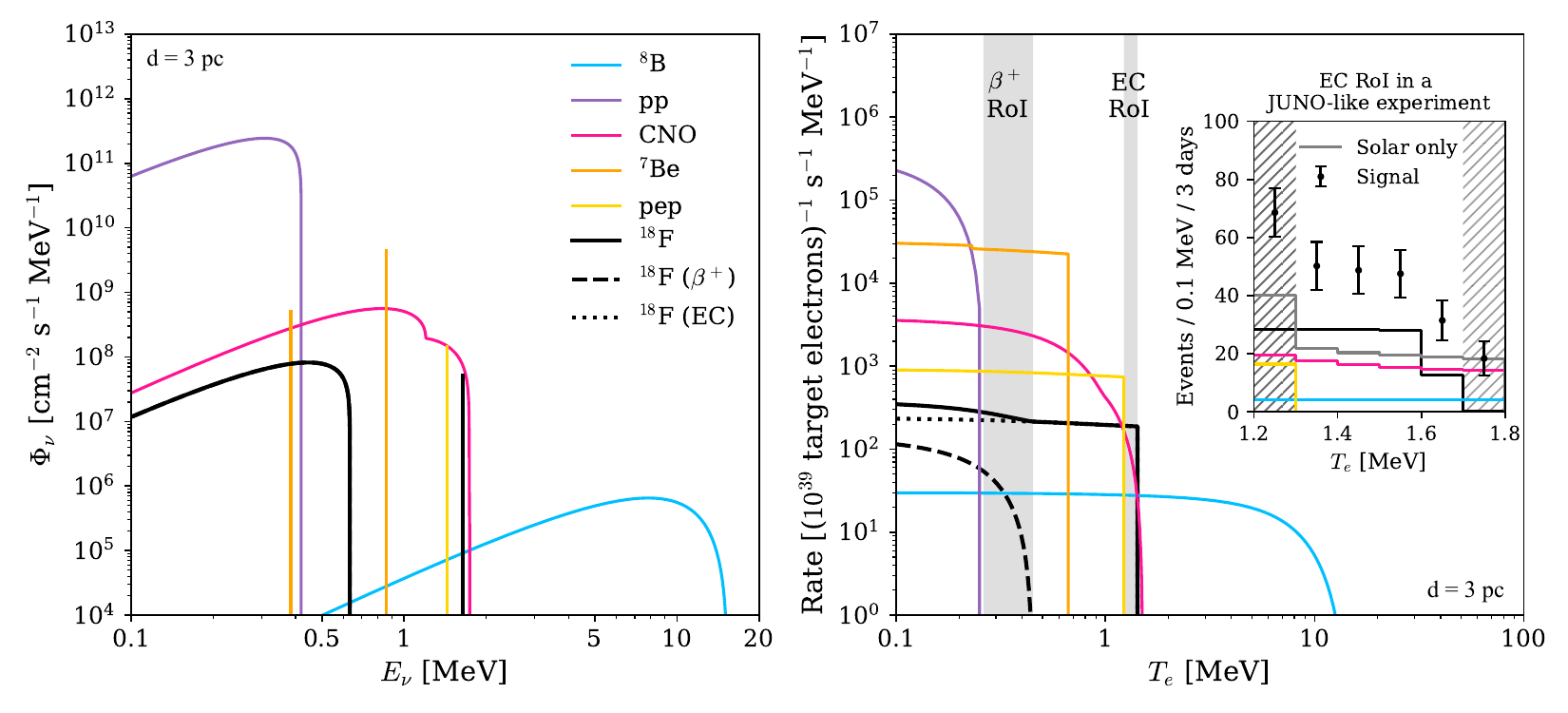}
    \caption{{\it Left panel:} All-flavor neutrino flux at Earth from $^{18}$F neutrinos  (black, thick curves)  and from the Sun (colored, thin curves)~\cite{Gonzalez-Garcia:2023kva}. The $^{18}$F fluxes are an average over the optimal $3.11$-day window (c.f.~Sec.~\ref{sec:window}) and a distance of $3$~pc. For line sources, flux units are ${\rm cm}^{-2}~{\rm s}^{-1}$. {\it Right panel:} Electron recoil spectrum (kinetic energy $T_e$) from elastic neutrino scattering, assuming standard flavor conversion in the Sun and  the  IO survival probability  in Eq.~\eqref{eq:IO-survival} for the $^{18}$F neutrinos. The gray vertical bands indicate the regions of interest (RoI) for the $\beta^+$ decay and EC source channels. The inset shows an example of the event rate in a JUNO-like experiment in the EC RoI.
    }
    \label{fig:flux_rate}
\end{figure*}

To develop an understanding of the $^{18}$F flux from a star at  $3$~pc from Earth, approximately representing the closest stars, we compare it in the left panel of Fig.~\ref{fig:flux_rate} with the different components of the  solar neutrino foreground~\cite{Gonzalez-Garcia:2023kva}. To be specific, we average the $^{18}$F flux over a $3.11$-day period determined in Sec.~\ref{sec:window}. These fluxes are also summarized in Table~\ref{tab:fluxes}.
 
The $^{18}$F flux spectrum  overlaps with the one of solar neutrinos. However, while the $\beta^+$ spectrum is buried under pp and CNO solar neutrinos, the EC line lies next to the tail of the CNO solar neutrino spectrum and thus may be easier to detect.

\begin{table}[]
    \centering
    \caption{All-flavor solar neutrino flux components~\cite{Gonzalez-Garcia:2023kva} and $^{18}$F neutrinos, averaged over the optimal window of $3.11$~days (c.f.~Sec.~\ref{sec:window}) and a source distance of $3$~pc. The neutrino energy for line sources and the maximum energy for continuum sources are also listed.
    \label{tab:fluxes} }
    \renewcommand{\arraystretch}{1.4}
    \begin{ruledtabular}
    \begin{tabular}{llll}  
    Process & $\Phi_{\rm tot}$ [cm$^{-2}$~s$^{-1}$] & E$_{\nu, \,\rm  max}$ [MeV] & E$_{\nu}$ [MeV]  \\
    \colrule
        pp & 5.94$\times10^{10}$     & 0.420  &\\
        \multirow{2}*{$^{7}$Be}& \multirow{2}*{$4.93\times 10 ^9$} & &0.862 (89.7\%) \\
        & & & 0.384 (10.3\%) \\
        pep & $1.42\times 10^{8}$ && 1.442\\
        $^8$B & $5.20\times 10^6$ & 15.2\\
        hep & $3.0\times10^4$&19.795\\
        $^{13}$C & $3.5\times 10^8$ &  1.199	\\
        $^{15}$N & $2.5\times 10^8$ & 1.732 \\
        $^{17}$F & $5.5\times 10^7$ & 1.740 \\ \colrule
        $^{18}$F ($\beta^+$) & 3.09$\times 10^7$ & 0.634 \\
        $^{18}$F (EC) & 5.10$\times 10^7$ & & 1.68
    \end{tabular}
    \end{ruledtabular}
\end{table}

\section{Neutrino detection prospects}
\label{sec:detection}

We now investigate the detection horizon for $^{18}$F neutrinos, assuming elastic electron scattering as the main channel in idealized low-energy neutrino observatories.

\subsection{Neutrino observatories}

The spectral flux expected from $^{18}$F neutrinos suggests that elastic electron scattering, notably of the EC line at $1.68$~MeV, is the most promising detection channel. Liquid scintillator detectors are suitable for the study of low-energy astrophysical neutrinos due to their low threshold for electron recoil. Borexino~\cite{BOREXINO:2023ygs}, in particular, succeeded in detailed solar neutrino spectroscopy and  measured the small CNO flux. JUNO~\cite{JUNO:2015zny} and SNO+~\cite{SNO:2015wyx} are operational and will continue that legacy. Further achievements could come from next-generation multi-purpose neutrino telescopes employing water-based liquid scintillators, such as the proposed Theia~\cite{Theia:2019non}, or located in low-background sites, such as the proposed Jinping neutrino experiment~\cite{Jinping:2016iiq}. We  here consider idealized versions of JUNO and the Jinping neutrino experiment as reference detectors.

\subsection{Detection process}

The most promising detection process is neutrino-electron elastic scattering, measuring the electron kinetic recoil energy $T_e$ in a scintillator. With a primary neutrino energy $E_\nu$, the maximum value is
\begin{equation}
    T_e^{\rm max}=\frac{2E_\nu}{2E_\nu+m_e}\,E_\nu\, ,
\end{equation}
with $m_e$ being the electron mass.
For our EC neutrino line with $E_\nu=1.68$~MeV, this is $T_e^{\rm max}=1.46$~MeV. The detectable electron spectrum is then essentially a flat spectral shoulder ending sharply at $T_e^{\rm max}$, as can be seen in the right panel of Fig.~\ref{fig:flux_rate}.

In terms of the inelasticity parameter $y=T_e/E_\nu$, the differential cross section is \cite{tHooft:1971ucy,Fukugita:2003en}
\begin{equation}
    \frac{d\sigma}{dy}=\sigma_0\left[c_L^2+c_R^2(1-y)^2-c_Lc_R\,\frac{m_e}{E_\nu}\,y\right]\, ,
\end{equation}
where in our case, $y_{\rm max}=0.868$. Moreover,
\begin{equation}
    \sigma_0=\frac{2G_{\rm F}^2E_\nu m_e}{\pi}=
    2.89\times10^{-44}~{\rm cm}^2\,\frac{E_\nu}{1.68~{\rm MeV}}\, ,
\end{equation}
with $G_{\rm F}=1.166\times10^{-5}~{\rm GeV}^{-2}$ being the Fermi constant. The coefficients are
\begin{equation}
    c_L=\pm\frac{1}{2}+\sin^2\Theta_W
    \quad\text{and}\quad
    c_R=\sin^2\Theta_W\, ,
\end{equation}
where the upper sign is for $\nu_e$, the lower one for $\nu_{\mu,\tau}$, and $\sin^2\Theta_W=0.2312$ \cite{ParticleDataGroup:2024cfk}.

For detecting the spectral shoulder in the electron spectrum, only a narrow energy range below $T_{e}^{\rm max}$ is relevant, as discussed later.  Therefore, the relevant cross sections are
\begin{subequations}
\begin{eqnarray}
   \frac{d\sigma_{\nu_e}}{dT_e}\Big|_{T_e^{\rm max}}&=& 8.45\times10^{-45}\,\frac{{\rm cm}^2}{{\rm MeV}}\, ,
   \\
   \frac{d\sigma_{\nu_{\mu,\tau}}}{dT_e}\Big|_{T_e^{\rm max}}&=& 1.54\times10^{-45}\, \frac{{\rm cm}^2}{{\rm MeV}}\, ,
\end{eqnarray}
\end{subequations}
with a ratio of $5.48$.

\subsection{Event rate}

The detection rate for neutrinos of  flavor $\nu_\alpha$ elastic scattering on electrons is, differential with regard to the electron kinetic energy $T_e$,
\begin{align}
    \frac{{\rm d}R_{i,\alpha}}{{\rm d}T_e} =N_e  \tau \int {\rm d}E_\nu  \frac{{\rm d}\Phi_{i,\alpha}}{{\rm d}E_\nu}\int {\rm d} T_e \frac{{\rm d}\sigma_\alpha (E_\nu)}{{\rm d}T_e}\, ;
\end{align}
here $i$ refers to the neutrino flux component (solar pp, $^7$Be, pep, $^8$B, hep, $^{13}$C, $^{15}$N, and $^{17}$F) as well as $^{18}$F neutrinos from $\beta^+$ decay and EC. Moreover, $E_\nu$ denotes the incoming-neutrino energy, $N_e$ the number of target electrons, and $\tau$ the  exposure time. 

The experimentally reconstructed $T_e$ value is not identical with the true value, but for a detector like JUNO, with its outstanding energy resolution, this effect is negligible for our exploratory study. At a visible energy of $2$~MeV, the energy resolution of JUNO is around $2\%$ \cite{JUNO:2024fdc} and roughly scales as $1/\sqrt{T_e}$. For our $E_\nu=1.68$~MeV and maximum recoil of $1.46$~MeV, the energy resolution is around $30$~keV, comparable to the line width of the EC line from the $^{18}$F source (see Appendix~\ref{sec:F18-decay}).

For the detector size, we are also guided by JUNO, which consists of $20$~kt liquid scintillator. Assuming a fiducial mass of $17$~kt, this  corresponds to $N_e = 5.7 \times 10^{33}$ target electrons. We assume $\tau=3$~days for the observation time (cf.~Sec.~\ref{sec:window}). In other words, we assume an exposure of $N_e\tau= 1.5\times 10^{39}$ electron-seconds. With these assumptions and ignoring backgrounds, Fig.~\ref{fig:flux_rate} (right panel) shows the event rate for $^{18}$F and solar neutrinos. We have here assumed standard solar flavor conversion, whereas for the $^{18}$F fluxes, we consider a $\nu_e$ survival probability corresponding to the fully adiabatic IO case of Eq.~\eqref{eq:IO-survival}. The elastic scattering cross section of $\nu_\mu$ or $\nu_\tau$ on electrons is around $18\%$ of the $\nu_e$ one so that $65\%$ of the signal comes from the surviving $\nu_e$ component.

The shape of the spectral rate allows us to identify two potential regions of interest (RoI). The first is the $\beta^+$~RoI with recoil energies in the range $0.26$--$0.45$~MeV, i.e.,~between the endpoint of the solar pp spectrum and the endpoint of the $^{18}$F spectrum. In this energy range, a steady signal from solar neutrinos is expected. The second is the EC RoI, defined by recoil energies $1.22$--$1.46$~MeV, the former being the maximum electron recoil energy induced by solar pep neutrinos and the shoulder caused by the EC neutrino line. Events in the EC RoI can also be induced by solar $^8$B and CNO neutrinos, specifically from $^{15}$O and $^{17}$F. As anticipated in Sec.~\ref{sec:nuemission},  any $^{18}$F detection opportunity is dominated by the EC line and its associated RoI.

\subsection{Detection horizon}
Opportunities for detecting our signal depend on the fiducial volume of existing and next-generation liquid-scintillation detectors and  their backgrounds (ignored in the right panel of Fig.~\ref{fig:flux_rate}). Relevant contributions in the RoI include  (i)~cosmic-ray muon induced spallation backgrounds, (ii)~noise due to the features of the active detector material regardless of the depth, or (iii)~environmental radioactive backgrounds.

A detection of the $^{18}$F neutrino burst with a certain significance would require that the number of signal events $N_{\rm EC}$ in the EC~RoI exceeds the statistical fluctuation of $N_{\rm B}$ background events by a certain factor. As a baseline background, we use the irreducible solar neutrino signal $N_\odot$ over a 3-day period plus $x$ times as much from other backgrounds. Hence, we define as a statistical measure
\begin{equation}
    \chi^2 = \frac{N_{\rm EC}^2}{(1+x)N_\odot}\, ,
\end{equation}
with $x=N_{\rm B}/N_\odot$. In the Gaussian limit of many events, a value of $\chi$ implies a $\chi$ sigma detection.

Besides $x$, the other crucial parameters are the source distance and the exposure $N_e\tau$. The event numbers for signal and background in the EC~RoI are found to be
\begin{subequations}
\begin{eqnarray}
    N_{\rm EC}&=& \begin{cases} 317\,(N_e\tau)_{39}/d_{\rm pc}^{2}\quad (\rm NO)\, ,
    \\
    645\,(N_e\tau)_{39}/d_{\rm pc}^{2}\quad \rm (IO)\, ,
    \end{cases}\\ \nonumber \\
    N_\odot&=& 85\,(N_e\tau)_{39}\, ,
\end{eqnarray}    
\end{subequations}
where $\tau=3$~days and we use $d_{\rm pc}=d/{\rm pc}$ for the source distance and $(N_e\tau)_{39}=N_e\tau/(10^{39}~\hbox{electron-seconds})$ for the exposure.

Therefore, a $\chi$ sigma detection requires that the parameters obey
\begin{equation}
    \chi< \frac{(N_e\tau)_{39}}{d_{\rm pc}^2\sqrt{1+x}}\times \begin{cases}
        34.4 \quad \hbox{(NO)}\, ,\\
        70.0 \quad \hbox{(IO)}\, .
    \end{cases}
\end{equation}
For an ideal detector ($x=0$), where the only background consists of solar neutrinos, with the same fiducial volume of JUNO, a nominal $\chi$ sigma detection would require a source distance of $d< 6.0 \,(8.6)/\sqrt{\chi}~{\rm pc}$ for NO (IO).

We stress that this significance only applies to a given 3-day observation window. However, as there is no other trigger than the neutrinos themselves, a search would require continuous observation over many years, during which large background fluctuations must statistically occur. In a ten-year period, a $3\sigma$ fluctuation within some 3-day window is practically certain to occur. However,  the significance could probably be improved by using more refined information on the neutrino pulse shape.

\begin{figure}
     \centering
     \includegraphics[width=\linewidth]{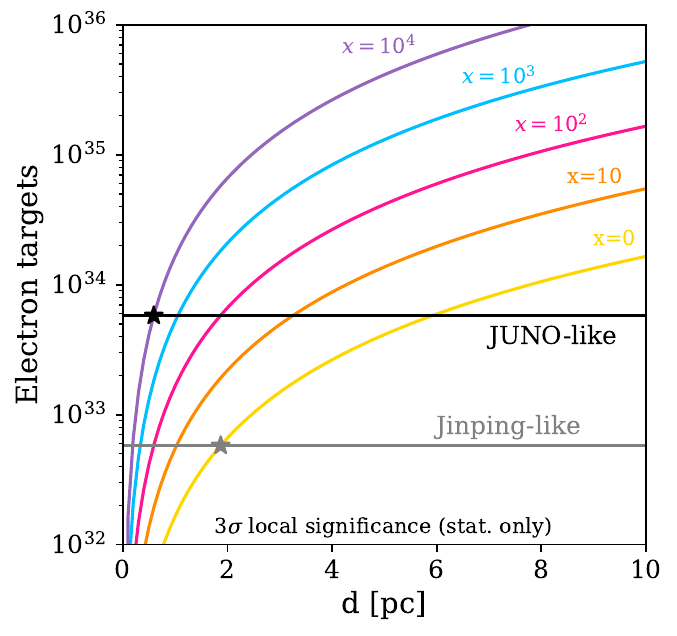}
     \caption{Number of electron targets in a liquid-scintillator neutrino telescope as a function of the source distance  to achieve a $3 \sigma$ (local significance)  detection of $^{18}$F for IO and accounting  for statistical uncertainties only (i.e., no systematic uncertainties). The curves of different colors represent  different assumptions on the background rate relative to the rate of solar neutrinos ($R_{\rm B} = x R_{\odot}$). The horizontal  lines indicate the number of target electrons in a neutrino telescope of size comparable to the one of JUNO~\cite{JUNO:2023zty} and Jinping experiment~\cite{Jinping:2016iiq}. An ideal background-free Jinping-like detector ($x=0$, cf.~gray asterisk)  would be sensitive to $^{18}$F neutrinos from sources up to $2.8$~pc with a $3\sigma$ local significance. A JUNO-like detector
     with such low background would be sensitive to $^{18}$F neutrinos from sources up to $5$~pc;  for a realistic JUNO value of $x\simeq 10^4$ (cf.~black asterisk), the source would need to be closer than $1$~pc.
     }
     \label{fig:distance}
 \end{figure}

Figure~\ref{fig:distance}  shows the same information in graphical form. It shows the number of target electrons $N_e$ as a function of the source distance such that $\chi > 3$ to assess the detection horizon and the experimental setup needed to achieve a $3\sigma$ detection when only statistical uncertainties are included. For more than a decade, Borexino took data in the EC~RoI. However, due to its small fiducial volume of only around $100$~ton~\cite{BOREXINO:2023ygs},  $N_e = 3.3\times 10^{31}$ and therefore we conclude from Fig.~\ref{fig:distance} that Borexino was not sensitive to $^{18}$F.

An ideal background-free  detector of the same size as JUNO (JUNO-like, horizontal line in Fig.~\ref{fig:distance}) would be sensitive to $^{18}$F neutrinos from sources  up to $5.9$~pc  at $3 \sigma$ for IO. For NO, the detection horizon at $3 \sigma$  is expected to shift to $4.2$~pc. However, the background rate of JUNO in the region of interest is approximately four orders of magnitude larger than the rate of solar neutrino events, i.e.,~$x \sim 10^4$~\cite{JUNO:2023zty}. Hence, the sensitivity of JUNO is limited to $\lesssim 1$~pc, hindering the discovery of $^{18}$F neutrinos independently of the neutrino mass ordering. Based on sensitivity studies for CNO solar neutrinos on a $20$~kton water based liquid scintillator like Theia~\cite{Bonventre:2018hyd}, one can infer that the sensitivity of Theia to $^{18}$F neutrinos would be comparable to the one of JUNO since they should have very similar $(N_e\tau)_{39}$. However, since the location and final volume of Theia are yet to be determined, the choice of a low-background site or a larger detector volume could improve the detection prospects.
  
An ideal detector of size comparable to the one of the Jinping neutrino experiment (Jinping-like, horizontal line in Fig.~\ref{fig:distance}, $(N_e\tau)_{39} = 0.15$ electron-seconds) would be sensitive to $^{18}$F neutrinos from sources up to $2.8$~pc away from us  for $3\sigma$ detections, for  IO. For NO,  a detection at $3\sigma$  may occur for sources at $\lesssim 2.0$~pc. Interestingly, the Jinping neutrino experiment is expected to have  a  background rate $x \lesssim \mathcal{O}(1)$~\cite{Jinping:2016iiq}; this implies that it would perform similarly to the idealized  detector considered~in~Fig.~\ref{fig:distance}.  
  
A similar background level [$x \lesssim \mathcal{O}(1)$] is expected for the EC RoI  in the liquid phase of SNO+~\cite{SNO:2021xpa}, although the latter has a fiducial volume approximately a factor $2$--$3$ smaller. Given its smaller volume, the sensitivity horizon of SNO+ would be accordingly reduced, hampering a significant detection. 
 
These findings highlight that EC neutrinos from $^{18}$F may be within reach with state-of-the-art technology. Although we expect a few He flashes per year in the Milky Way (see Appendix~\ref{sec:He-flash-rate}), to date the closest known source candidate is Arcturus at  $11.3$~pc from Earth. Hence, we conclude that known source candidates are  beyond the detection horizon of existing and upcoming neutrino telescopes. The detection chances of the  neutrino spectrum from $^{18}$F $\beta^+$ decay are worse than the ones for the EC channel, in agreement with the early findings of Ref.~\cite{Serenelli:2005nh}.

We stress that we investigate idealized detection prospects to assess the features that neutrino telescopes should have to be sensitive to $^{18}$F neutrinos. In order to do so and because of our poor understanding of the detector backgrounds in the EC RoI, we focused on energy integrated quantities in Fig.~\ref{fig:distance}. However, spectral information may be crucial in reducing the impact of detector backgrounds. Moreover, directional information could also play a role in reducing isotropic and solar backgrounds.

\section{Conclusions}
\label{sec:conclusions}

The burst of $^{18}$F neutrinos caused by the He flash in low-mass stars is one of the brightest episodes of neutrino emission in stellar evolution, as first shown by Serenelli and Fukugita \cite{Serenelli:2005nh}. Only the most advanced burning stages of massive stars provide larger fluxes \cite{Farag:2020nll, Martinez-Mirave:2025dae}, and of course core-collapse supernovae \cite{Mirizzi:2015eza, Raffelt:2025wty,Tamborra:2024fcd}. In this paper, we have re-examined the neutrino emission from $^{18}$F decay. For the first time, we have identified the importance of electron capture on $^{18}$F that provides an additional monoenergetic line at $1.7$~MeV. We have explored the detection prospects for these $^{18}$F neutrinos in existing and near-future neutrino observatories.

We find that the detection of $^{18}$F neutrinos in JUNO is hampered by natural detector backgrounds. However, the upcoming Jinping neutrino experiment could ideally detect this signal from stars up to $3$~pc at $3 \sigma$, owing to its very low backgrounds in the energy region of interest.

Existing technology is well suited for the detection of $^{18}$F neutrinos in sites with low natural backgrounds. However, unfortunately, the $^{18}$F detection prospects are dim due to the lack of target stars within the detection distance horizon: the closest conceivable possibility is the red-giant-branch star Arcturus at a distance of $11.3$~pc. Therefore, as of now, asteroseismology remains the best bet to test our understanding of the inner workings of evolved low-mass stars.

\section*{Acknowledgments}
We  thank Achim Weiss for illuminating discussions. This project has received support from the Villum Foundation (Project No.~13164), the Elite Research Prize  from the Danish Ministry of Higher Education and \hbox{Science} (Project No.~3142-00074B), and the German Research Foundation (DFG) through the Collaborative Research Center ``Neutrinos and Dark Matter in Astro- and Particle Physics (NDM),'' Grant No.\ SFB-1258-283604770. We used the Tycho supercomputer hosted at the SCIENCE HPC center at the University of Copenhagen to support our numerical simulations.

\appendix

\section{Beta processes \texorpdfstring{\boldmath{$^{18}$}F$\to$\boldmath{$^{18}$}O}{}}
\label{sec:F18-decay}
In this appendix, we provide additional details on the beta decay and EC processes linked to the decay of ${}^{18}{\rm F}$ to ${}^{18}{\rm O}$. We then contrast our findings with the tabulated rates from Ref.~\cite{Oda:1994} adopted in \texttt{MESA}.

\subsection{Beta decay}

The weak decay ${}^{18}{\rm F}\to{}^{18}{\rm O}+e^++\nu_e$ is of the type $1^+\to 0^+$, therefore an allowed pure Gamow-Teller transition. The $Q$ value is $0.6339$~MeV \cite{NuclearData,Tilley:1995zz}, which is the maximum neutrino energy or the maximum positron kinetic energy.  The maximum positron total energy, which is also the mass difference between the two nuclei, is
\begin{equation}
  E_0=Q+m_e=1.1449~{\rm MeV}\, ,
\end{equation}
with $m_e$ being the positron mass.

The first excited state of ${}^{18}{\rm F}$ is at $0.9372$~MeV and it does not play a role for He-flash conditions. The first excited state of ${}^{18}{\rm O}$ is at $1.9821$~MeV that is considerably higher than the ground state of ${}^{18}{\rm F}$, i.e., the $\beta^+$ decay or EC is only from ground state to ground state.

The measured half-life is $1.82871(18)$~h \cite{GarciaTorano:2010}, with  branching ratio of $96.86(19)\%$ for the $\beta^+$ channel and $3.14(19)\%$ for atomic EC. Therefore, 
\begin{equation}\label{eq:Gamma-F18}
  \lambda_{\beta+}=1.020 \times10^{-4}~{\rm s}^{-1}=6.714 \times 10^{-20}~{\rm eV}
\end{equation}
is the rate for the $\beta^+$ channel.

The rate of such a decay, differential with regard to electron and neutrino momentum ($p_e$ and $p_\nu$, respectively), integrated or averaged over spin and angle and  neglecting nuclear recoil, is given by:
\begin{equation}\label{eq:decay-rate}
  d\lambda_{\beta^+}=\lambda_0\, \frac{30}{E_0^5}\,p_e^2dp_e p_\nu^2dp_\nu\,F(Z,E_e)\,\delta(E_0-E_e-E_\nu)\, ,
\end{equation}
where $\lambda_0$ is a constant that involves all coupling constants and the Gamow-Teller nuclear matrix element, $E_e=\sqrt{p_e^2+m_e^2}$ is the electron energy, $E_\nu$ is the neutrino energy, and $F(Z,E_e)$ is the Fermi function accounting for Coulomb corrections, with $Z$ being the atomic number. For $\beta^+$ decay, $Z$ is the negative of the daughter nuclear charge;  in the $^{18}$F case, $Z=-8$. The numerical prefactor equal to $30$ in Eq.~(\ref{eq:decay-rate}) has been chosen so that the integral over phase space yields $\lambda_{\beta^+}=\lambda_0$, when neglecting Coulomb corrections and in the limit $m_e\to 0$.

If we ignore screening effects in the plasma, the Fermi function is \cite{Fuller:1980zz}
\begin{equation}
  F(Z,E_e)=2(1+s)(2p_eR)^{2s-2} e^{\pi\eta}
  \left|\frac{\Gamma(s+i\eta)}{\Gamma(2s+1)}\right|^2\, ,
  \label{eq:fermifunction}
\end{equation}
where $p_e=\sqrt{E_e^2-m_e^2}$,
$\eta=\alpha Z E_e/p_e=\alpha Z/v_e$ is the Sommerfeld parameter, $\alpha$ is the fine-structure constant, $s=\sqrt{1-(\alpha Z)^2}$, and $R$ is the nuclear radius. A simple estimate is $R=A^{1/3}\,1.2~{\rm fm}$.
For our relatively small-$Z$ nuclei, a common approximation is
\begin{equation}
  F(Z,E_e)\simeq \frac{2\pi\eta}{1-e^{-2\pi\eta}}\, ,
\end{equation}
which is accurate within a few percent.

Integrating over the electron phase space provides the neutrino spectrum for allowed Gamow-Teller transitions:
\begin{equation}
  \frac{d\lambda_{\beta^+}}{dE_\nu}=\lambda_0\,\frac{30\,E_\nu^2 E_ep_e}{E_0^5}\,F(Z,E_e)\, ,
\end{equation}
where $E_e=E_0-E_\nu$ and $p_e=\sqrt{E_e^2-m_e^2}$. Integrating over $E_\nu$ yields
\begin{equation}
  \lambda_{\beta^+}=0.289\,\lambda_0\, ,
\end{equation}
which includes a $24\%$ reduction due to Coulomb corrections. The average energy without Coulomb corrections would be $0.393$~MeV, whereas including the Fermi function yields $0.384$~MeV.

\subsection{Electron capture}

As for  EC, we need to put the charged lepton in the initial state and multiply with the occupation number. Hence  Eq.~\eqref{eq:decay-rate} becomes
\begin{eqnarray}\label{eq:capture-rate}
  d\lambda_{\rm EC}&=&\lambda_0\, \frac{30}{E_0^5}\,p_e^2dp_e p_\nu^2dp_\nu\,\delta(E_0+E_e-E_\nu)
  \nonumber\\
  &&\kern3em{}\times f_e(p_e)\,F(Z,E_e)\, ,
\end{eqnarray}
with the Fermi-Dirac distribution $f_e=[e^{(E_e-\mu_e)/T}+1]^{-1}$ in terms of the electron chemical potential $\mu_e$. The differential neutrino spectrum is then
\begin{equation}\label{eq:EC-spectrum}
  \frac{d\lambda_{\rm EC}}{dE_\nu}=\lambda_0\,\frac{30\,E_\nu^2 E_e^2}{E_0^5}
   \,\frac{v_eF(Z,E_e)}{e^{(E_e-\mu_e)/T}+1}
\end{equation}
with $E_e=E_\nu-E_0$, $p_e=\sqrt{E_e^2-m_e^2}$, and
$v_e=p_e/E_e$. 

For EC, the nuclear charge is the one of the initial-state nucleus, in our case $Z=+9$. Here, the electron wave function is enhanced at the location of the nucleus, leading to an increase in the rate, instead of a suppression as in the $\beta^+$  decay. Notice that near threshold for non-relativistic electrons, $F(Z,E_e)$ diverges as $1/v_e$, whereas $G(Z,E_e)=v_e F(Z,E_e)\simeq 2\pi\alpha Z$ is a constant for $v_e\to 0$.

When the electrons are completely degenerate, so that $\mu_e$ is identical with the Fermi energy $E_{\rm F}$, the maximum neutrino energy is $E_\nu^{\rm max}=E_0+E_{\rm F}$. If there are two nucleons per electron, the Fermi momentum in terms of the mass density is
\begin{equation}
  p_{\rm F}=0.190~{\rm MeV}\,\left(\frac{\rho}{10^5~{\rm g}/{\rm cm}^3}\right)^{1/3}\, .
\end{equation}
For a typical He-flash density of $2\times10^5~{\rm g}~{\rm cm}^{-3}$, this is about $p_{\rm F}=0.239$~MeV, which we use as a reference value, with $E_{\rm F}=0.564$~MeV. With these parameters, and without the Fermi function, the EC rate would be $\lambda_{\rm EC}=0.197\,\lambda_0$, whereas including it, one finds
\begin{equation}
  \lambda_{\rm EC}=0.381\,\lambda_0\, ,
\end{equation}
and thus an enhancement by a factor of $1.93$. For these conditions, EC capture exceeds the $\beta+$ neutrino production rate in that $\lambda_{\rm EC}/\lambda_{\beta^+}=1.32$. The two rates would be approximately equal for $\rho=1.5\times10^5~{\rm g}~{\rm cm}^{-3}$.

In the case of atomic EC, the line has an energy $E_\nu=E_0+m_e=1.656$~MeV. For our reference conditions, instead, the average energy is $\langle E_\nu\rangle=1.685$~MeV and thus around $30$~keV higher. For smaller electron densities and thus smaller electron kinetic energies, this excess is smaller. The range of energies is then from $E_0+m_e<E_\nu<E_0+E_{\rm F}$; hence, the line has a width of $53$~keV.

The spectral shape of the EC line given by Eq.~\eqref{eq:EC-spectrum} is approximately $\lambda_0 30 (E_0+m_e)^2 m_e^2/E_0^5$ times a slowly varying function of $v_e$, which is $2\pi\alpha Z$ at threshold. For complete degeneracy and nonrelativistic electrons, this is essentially a top-hat shape.

For non-degenerate conditions and $T=10$~keV, the average electron kinetic energy is still $15$~keV. So the line is shifted by about $15$~keV beyond its nominal energy. Thermal motions of the nuclei and gravitational redshift, by comparison, are much smaller effects. The refractive modification of the electron mass  is  also too small to~matter.

\subsection{Tabulated rates}

The \texttt{MESA} code does not evaluate the beta-decay and EC formulas derived in this appendix, instead it uses the tabulated rates from Oda et al.\ \cite{Oda:1994}. These are given for values of $\log_{10}(Y_e\rho)$ in steps of $1$ (i.e., one value per order of magnitude, and also a crude grid of temperatures). At $T$ so low that there are no positrons in the medium, the rate for the $\beta^+$ decay of $^{18}$F is reported as $\lambda_{\beta^+}=10^{-4.011}~{\rm s}^{-1}=0.97\times10^{-4}~{\rm s}^{-1}$, to be compared with Eq.~\eqref{eq:Gamma-F18}, i.e., about $5\%$ smaller. From the description in Ref.~\cite{Oda:1994}, it is not clear if  the measured or the calculated lifetime was used for this particular nucleus, therefore the origin of this small discrepancy is not obvious.

For EC capture, our reference density corresponds to $Y_e\rho=10^5~{\rm g}~{\rm cm}^{-3}$ so that we may compare to the tabulated value at this point on the grid and at the lowest tabulated $T=10^7$~K, which assures complete degeneracy. Reference~\cite{Oda:1994} finds $\lambda_{\rm EC}=10^{-3.889}~{\rm s}^{-1}$, similar to the $\beta^+$ rate. Their ratio is $\lambda_{\rm EC}/\lambda_{\beta^+}=10^{-3.889+4.011}= 1.32$, in  agreement with what we have found.

\section{Helium flash rate in the Galaxy}

\label{sec:He-flash-rate}

In Ref.~\cite{Martinez-Mirave:2025dae}, the neutrino emission from all stars in the Milky Way was computed, including an average flux of $^{18}$F neutrinos. However, the transient contribution from the He flash was not separated from the steady neutrino emission expected throughout the life of stars of different masses.  We here isolate the  He-flash contribution more explicitly to infer the likelihood of such  events in the Milky Way.

The stellar population of the Galaxy is modeled in terms of a stellar birth rate function, namely
\begin{align}
    B(t,m)= {\rm SFH}(t)\, {\rm IMF}(m)\, ,
\end{align}
where SFH is the star formation history, $t$ is the cosmic time, IMF is the initial mass function, and $m = M/M_\odot$.

The main-sequence lifetime of stars in the mass range of interest is much larger than the time between the end of hydrogen burning in the core and the He flash. Then, to a good approximation, the age at which the He flash occurs can be related to the ZAMS mass as
\begin{align}
    \tau_{\rm He} \sim 10 \left(m\right)^{\alpha} {\rm Gyr}\, ,
\end{align}
where we have calibrated $\alpha = -3.5$ with our suite of $\texttt{MESA}$ models in the mass range of interest and it roughly agrees with the results of Ref.~\cite{2009itss.book.....P}. We have also checked that the choice of $\alpha$ does not have a strong impact on the following result. 

Given the present number of stars in the Galaxy ($N_*$), the number of expected He flashes per year is given by 
\begin{align}
   R_{\rm He \, flash} =
    N_*&\int_{m_{\rm min}}^{2.2} {\rm SFH}[\tau_{\rm U} - \tau_{\rm He}(m)]\nonumber \\  &\times{\rm IMF}(m) \sqrt{1 - \left(\frac{{\rm d}\tau_{\rm He}}{{\rm d} m}\right)^2} {\rm d}m \, .
\end{align} 

Here, $m_{\rm min} = 0.917$ is the minimum stellar mass for which $\tau_{\rm He} > \tau_{U}$, with $\tau_U =13.5$~Gyr being the age of the Universe. Following this procedure separately for the bulge, the thin disk, and the thick disk, and adopting the parametric star formation history  and estimated number of stars from Ref.~\cite{Martinez-Mirave:2025dae}, we find: 
\begin{align}\label{eq:annual-He}
    R_{\rm He\, flash} = 12 \, {\rm yr}^{-1}
\end{align}
for the annual rate of He flashes in the Galaxy.

Given that the average number of neutrinos emitted over a year after the He flash is $2\times 10^{45}$ s$^{-1}$,  Eq.~\eqref{eq:annual-He}, and assuming an average source distance of $8$~kpc (based on the matter distribution of the Milky Way), one finds an expected flux at Earth of $3$~cm$^{-2}$~s$^{-1}$. This result is one order of magnitude smaller than $30.2$~cm$^{-2}$~s$^{-1}$ quoted in Ref.~\cite{Martinez-Mirave:2025dae} for all neutrinos from $^{18}$F. In fact, the latter also accounted for the long-term neutrino emission after the He flash, as well as the contributions from He subflashes and thermal pulses.

Notice that in Ref.~\cite{Martinez-Mirave:2025dae}, the flux from $^{18}$F is presented as a steady flux. However, this is not relevant when addressing a hypothetical detection since longer time exposures of the order of (at least) years would be needed. Moreover,  the spectral shape of the corresponding flux would feature the continuum spectrum from $\beta^+$ decay and the EC emission line.

\bibliography{bibliography}

\end{document}